\documentclass[letterpaper, 10 pt, conference]{ieeeconf}  

\IEEEoverridecommandlockouts                              

\usepackage{cite}
\usepackage{amsmath,amssymb,amsfonts}
\usepackage{algorithmic}
\usepackage{graphicx}
\usepackage{textcomp}

\usepackage{array}
\usepackage{graphicx}
\usepackage{mathtools}
\usepackage{hyperref}
\usepackage{makecell}

\usepackage{caption}
\usepackage{subcaption}
\usepackage[font=small,labelfont=bf]{caption}

\usepackage{float}
\floatstyle{plaintop}
\restylefloat{table}
\usepackage{gensymb}

\DeclareMathOperator*{\argmin}{arg\,min}

\def\BibTeX{{\rm B\kern-.05em{\sc i\kern-.025em b}\kern-.08em
    T\kern-.1667em\lower.7ex\hbox{E}\kern-.125emX}}

\begin{document}
\title{\LARGE \bf
Localizing Scan Targets from Human Pose for \\Autonomous Lung Ultrasound Imaging
}
\author{Jianzhi Long$^{1,2}$, Jicang Cai$^{2}$, Abdullah Al-Battal$^{2}$, Shiwei Jin$^{2}$, \\ Jing Zhang$^{1}$, Dacheng Tao$^{1,3}$, 
Imanuel Lerman$^{2}$, 
Truong Nguyen$^{2}$
\thanks{$^{1,2}$J. Long is with the School of Computer Science, The University of Sydney, NSW 2006 AU. (e-mail: jianzhi199@gmail.com), this work is partially done during his master program at University of California San Diego.}
\thanks{$^{2}$J. Cai, A. Al-Battal, S. Jin, I. Lerman, T. Nguyen are with the Department of Electrical and Computer Engineering, University of California San Diego, CA, 92161 USA. (e-mail: jianzhi199@gmail.com, j1cai@ucsd.edu, aalbatta@eng.ucsd.edu, sjin@eng.ucsd.edu, ilerman@health.ucsd.edu, tqn001@eng.ucsd.edu)}
\thanks{$^{1}$J. Zhang and D. Tao is with the School of Computer Science, The University of Sydney, NSW 2006 AU. (e-mail: jing.zhang1@sydney.edu.au, dacheng.tao@gmail.com)}
\thanks{$^{3}$D. Tao is also with the JD Explore Academy, Beijing, CN.}
}

\maketitle
\thispagestyle{empty}
\pagestyle{empty}

\begin{abstract}

Ultrasound is progressing toward becoming an affordable and versatile solution to medical imaging. With the advent of COVID-19 global pandemic, there is a need to fully automate ultrasound imaging as it requires trained operators in close proximity to patients for a long period of time, therefore increasing risk of infection. In this work, we investigate the important yet seldom-studied problem of scan target localization, under the setting of lung ultrasound imaging. We propose a purely vision-based, data driven method that incorporates learning-based computer vision techniques. We combine a human pose estimation model with a specially designed regression model to predict the lung ultrasound scan targets, and deploy multiview stereo vision to enhance the consistency of 3D target localization. While related works mostly focus on phantom experiments, we collect data from 30 human subjects for testing. Our method attains an accuracy level of $16.00 \pm 9.79$mm for probe positioning and $4.44 \pm 3.75$\degree for probe orientation, with a success rate above $80\%$ under an error threshold of 25mm for all scan targets. Moreover, our approach can serve as a general solution to other types of ultrasound modalities. The \href{https://github.com/JamesLong199/Autonomous-Transducer-Project}{code} for implementation has been released.

\end{abstract}


\section{Introduction}
\label{sec:introduction}
Ultrasound (US) imaging has been increasingly prevalent in modern day clinical practice. Nowadays, the scope of US imaging modalities has covered almost all locations on the body, including brain, thyroid, heart, breast, fetal and prostate \cite{avola2021ultrasound}. Compared to other alternatives like Computed Tomography (CT) and Magnetic Resonance Imaging (MRI), US imaging is harmless, cost effective, portable and can provide real-time feedback \cite{li2021overview}. One major drawback for US imaging, though, is its dependence on the expertise of human sonographers to produce quality images, which is the main reason for the numerous efforts by researchers to develop Robotic Ultrasound Systems (RUS) \cite{priester2013robotic}. However, many of these robotic systems either provide aid to freehand scanning, or require an operator to input instructions \cite{li2021overview}. With the advent of the COVID-19 global pandemic, it is necessary to completely eliminate the risk of infection between patients and medical staff, especially for lung US scan. Therefore, a fully autonomous US imaging system will ensure both consistency in US image quality and safety for medical practitioners. 

Though existing autonomous US imaging system are designed for different types of US modality, most of these systems still share a considerable degree of similarities. To be specific, first, a scan-path planning algorithm defines a set of waypoints for the US probe, based on data from sensors like RGB-D cameras. A position and force control module is in place for adjusting the US probe, to ensure optimal contact for acquiring high quality US images and the safety of the subject. Finally, US image processing techniques are applied to enhance image quality for further evaluation. There are many works that address position and force control of US by fine-tuning the position and orientation of US probe, usually through feedback from force sensors \cite{li2021overview}. Moreover, a substantial amount of effort has been invested in the topic of US image processing, including image segmentation, pathology detection and classification \cite{avola2021ultrasound}, as well as 3D volume reconstruction \cite{mohamed2019survey}. 

In comparison, there are very few works on automatic scan path planning. It typically consists of two steps: first, the transducer probe is moved to the proximity of the target scan location, and then a scan path can be constructed in the neighboring region of the starting point. The first step, known as the scan target localization problem \cite{ma2021autonomous}, is particularly crucial for achieving truly autonomous US scan and yet still very much underexplored. The transducer probe needs to be placed in the close proximity of the starting point of the scan in the first place in order to initiate the following scanning process. Until now, there are only a few works that address this problem \cite{huang2018fully, lee2018automated, yang2021automatic, ma2021autonomous}. The reason behind this lack of research effort might be the lack of relevant data in comparison to the wide range of US imaging tasks and the variability in human anatomy among patients. So far, even though existing works have explored a variety of methods, none of them could be readily adopted as a general US imaging solution for human subjects.

In this paper, we introduce a scan target localization system for autonomous lung ultrasound imaging, which includes an ultrasound machine, two RGB-D cameras, a 6-DOF industrial robot, and a computer that controls the visual sensors and the robot. We assume that the image of human torso can provide rich information on the location of human internal organs, and consider US positioning as a variant to the existing human pose estimation problem in computer vision \cite{chen2020monocular}. We propose a target localization algorithm that learns from visual data, which integrates a pose estimation model with a scan target regression model with empirically determined parameters. Multiview Stereo Vision (MVS) is another crucial component in our system, for accurately recovering the 3D information of the features extracted from 2D images.

 We evaluate our system on 30 human subjects with diverse body composition. We also discuss the potential of our method as a general and readily applicable solution to other imaging targets for US scanning. 
In summary, our contribution is three-fold and can be summarized as follows: 

\begin{itemize}
  \item We present a vision-based autonomous lung US scan target localization system, which could also be generalized to other imaging modalities.
  \item We approach the scan target localization problem as a variant of the human pose estimation problem in computer vision.
  \item We collect data from and test our system on 30 human subjects, while related works mostly involves significantly fewer subjects or entirely focus on phantom experiments.

\end{itemize}

\section{Related Works}

We consider our work as an interdisciplinary application of computer vision in the field of robotic medical image acquisition. The core of our system is a target localization algorithm based on human pose estimation. We extend the model for traditional human pose estimation problem to locating US scan targets on the human body.

Human Pose Estimation (HPE) is widely used in areas ranging from virtual reality to medical assistance. It aims to obtain posture of the human body from images or videos \cite{chen2020monocular}. Given an input image that contains a human, the algorithm outputs the image pixel coordinates of the detected body part keypoints. In recent years, deep learning has facilitated rapid progress in this field. Deep learning based HPE algorithms are divided into two main categories: the top-down approach and the bottom-up approach. The top-down model consists of a human body region detector and a single person pose estimator \cite{xu2022vitpose}. The detector outputs bounding boxes surrounding each human subject in the image. The pose estimator is then run through each cropped bounding box to obtain the corresponding keypoint locations. On the other hand, the main components of most bottom-up methods include body joint detection and joint candidate grouping \cite{cao2017realtime}. The algorithm first predicts all the 2D joints present in the image and then assembles them into independent skeletons. In comparison, the runtime of top-down methods is often much slower, and has a linear relationship with the number of people detected in the image. However, the top-down methods usually yield better results, achieving state-of-the-art performance on most benchmarks.

The existing vision-based US scan target localization methods can be divided into two categories: non-learning based and learning based. For non-learning based methods, Huang et al. \cite{huang2018fully} locate the scan region by extracting the contour of phantoms with a rule-based image segmentation method, which is not applicable to real human subjects as the human body is continuous and cannot be physically divided into separate sections. Lee et al. \cite{lee2018automated} rely on nipples as the primary feature to determine the target for breast US scan with SURF feature extraction algorithm \cite{bay2006surf}. This method lacks generalization ability to other imaging targets. Tan et al. \cite{tan2022fully} collect chest shape measurements from 331 volunteers, and derive boundaries of a fixed 3D space that contains the chest region of every subject, which guides a rule-based path generation algorithm. This method lacks flexibility and generalization ability in terms of experiment setup. 

There are three different approaches for learning-based scan target localization methods. Focusing on US scan for human spine, Yang et al. \cite{yang2021automatic} develop an end-to-end deep learning based model that directly extracts the segmentation map of the entire longitudinal human spine area from the color and depth data. Although possibly applicable to knee joint and thyroid imaging, it could be challenging for this approach to adapt to more spread-out and deformable areas on human body such as breast and abdomen. A Reinforcement Learning (RL) based method has been proposed by Ning et al. \cite{ning2021autonomic}, which directly computes robot action based on multimodal data including 2D scene image, force measurement and US images. This method could become a general solution in the future, however currently it is still a concept model for phantom experiments. Lastly, Ma et al. \cite{ma2021autonomous} provide an HPE based approach for lung US. They assume the UV texture coordinate of each scan target is constant for all individual subjects, and employ a DensePose model \cite{guler2018densepose} to find the mapping between the UV texture and image pixel coordinates of the scan targets. Besides the constraining assumption, the prospect of general application of this method needs further verification as all the experiments are conducted on a single mannequin.


\section{Methods}%

\subsection{Target Scan Locations}
\begin{figure}[ht]
    \centering
    \includegraphics[width=6.5cm]{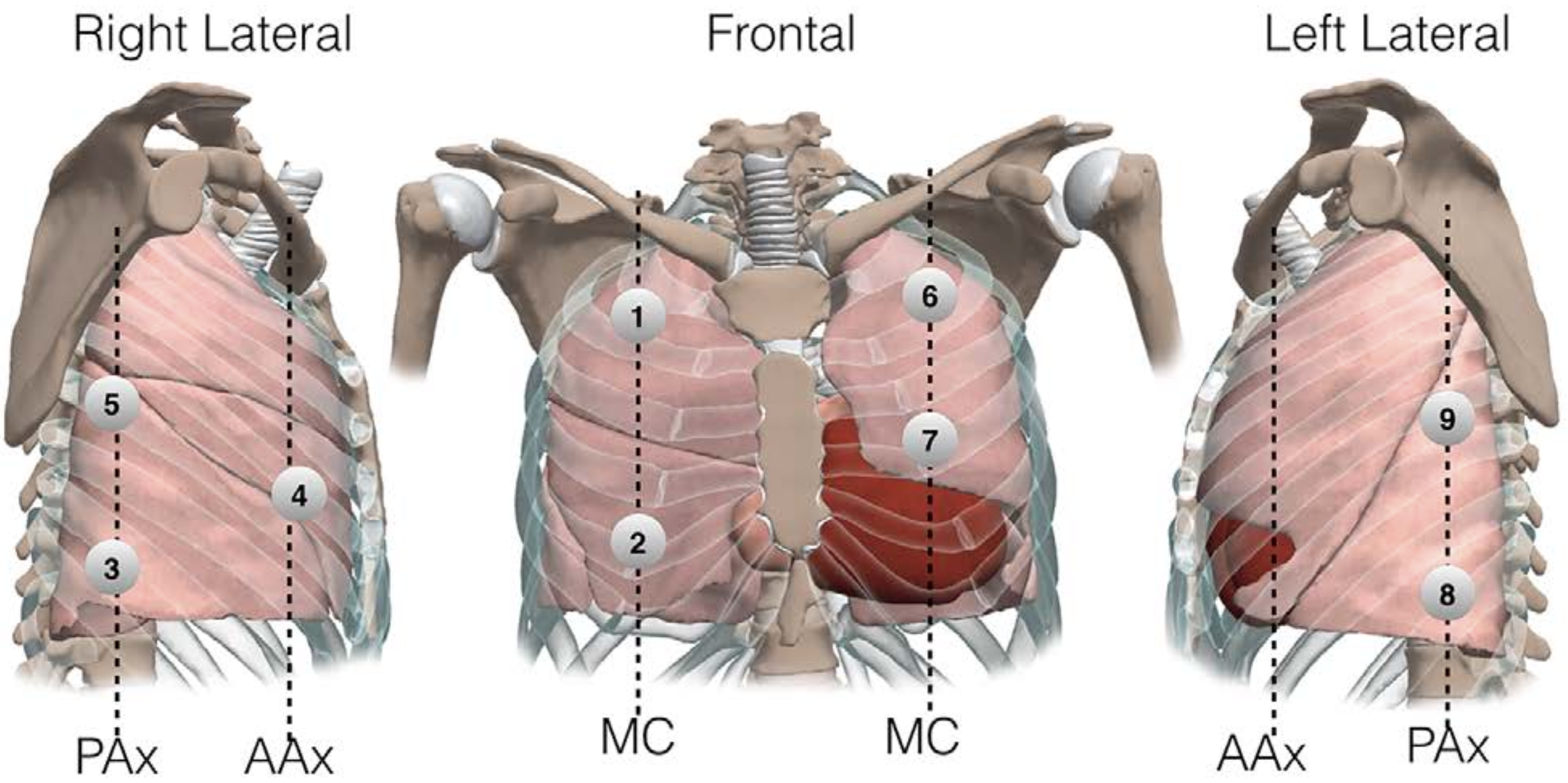}
    \caption{9 Target Scan Locations. AAx = anterior axillary line, PAx = posterior axillary line, MC = mid-clavicular line (image adopted from \cite{tierney2020comparative}). We focus on Target 1, 2, 4.}
    \label{target_scan_locations}
\end{figure}

Fig. \ref{target_scan_locations} shows the target scan locations according to the 9-point pulmonary ultrasound protocol specified by Tierney et al.\cite{tierney2020comparative}. The nine locations include front and lateral points, and are spread across both the left and right side of the human body. Due to the limited reach of our robot arm, we focus on scan locations on the right side of the torso. Moreover, performing US scan on the Posterior Axillary Line (PAx) would be inconvenient for all patients, and infeasible for those relying on ventilators. As a result, we test our method on scan locations 1, 2 and 4, situated on the Mid-Clavicular Line (MC) and Anterior Axillary Line (AAx). 

\subsection{System Design}
\begin{figure}[ht]
    \centering
    \includegraphics[width=6.5cm]{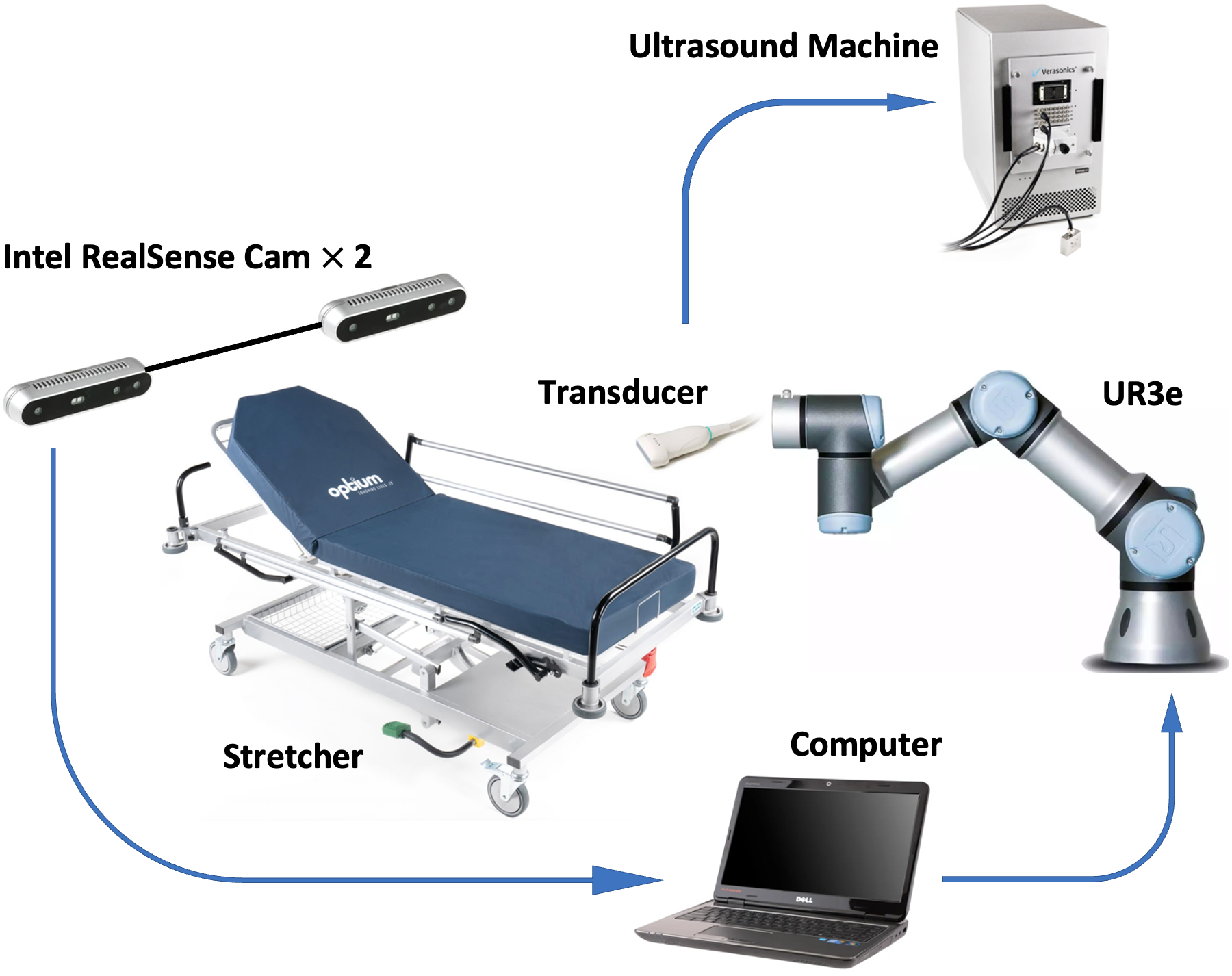}
    \caption{System components and setup. The arrows indicate the direction which information travels. The computer sends instruction to the robot based on camera data. The output from the ultrasound machine is not used in our computation pipeline.}
    \label{Apparatus}
\end{figure}

Our system consists of the following components: an industrial 6-DoF robot arm (Universal Robots UR3e), a Verasonics US machine paired with a Verasonics L12-3V linear array transducer, two Intel RealSense D-415 cameras, a PC (Dell Inspiron 16 Plus) that controls the visual sensors and robot, a camera stand and a stretcher (see Fig. \ref{Apparatus}). We use a UR python library developed by Rope Robotics (Denmark) to program and manipulate the robot arm, with an Ethernet cable connecting the PC and the robot. The transducer probe is mounted on the robot arm gripper with a customized holder. The two RGB-D cameras are mounted on the camera stand to capture images of the subject lying on the stretcher. The RGB-D cameras have an ideal sensing range from 0.5m to 3m, and their resolution is set to 640 $\times$ 480 pixels. The target scan poses are computed from the captured two-view RGB-D images, to which the robot arm is programmed to move. Our scan target localization algorithm does not rely on ultrasound image, therefore in practice our system does not process the ultrasound machine output. 

\subsection{Pipeline}

\begin{figure*}[ht]
    \centering
    \begin{subfigure}[b]{0.64\textwidth}
        \centering
        \includegraphics[width=\textwidth]{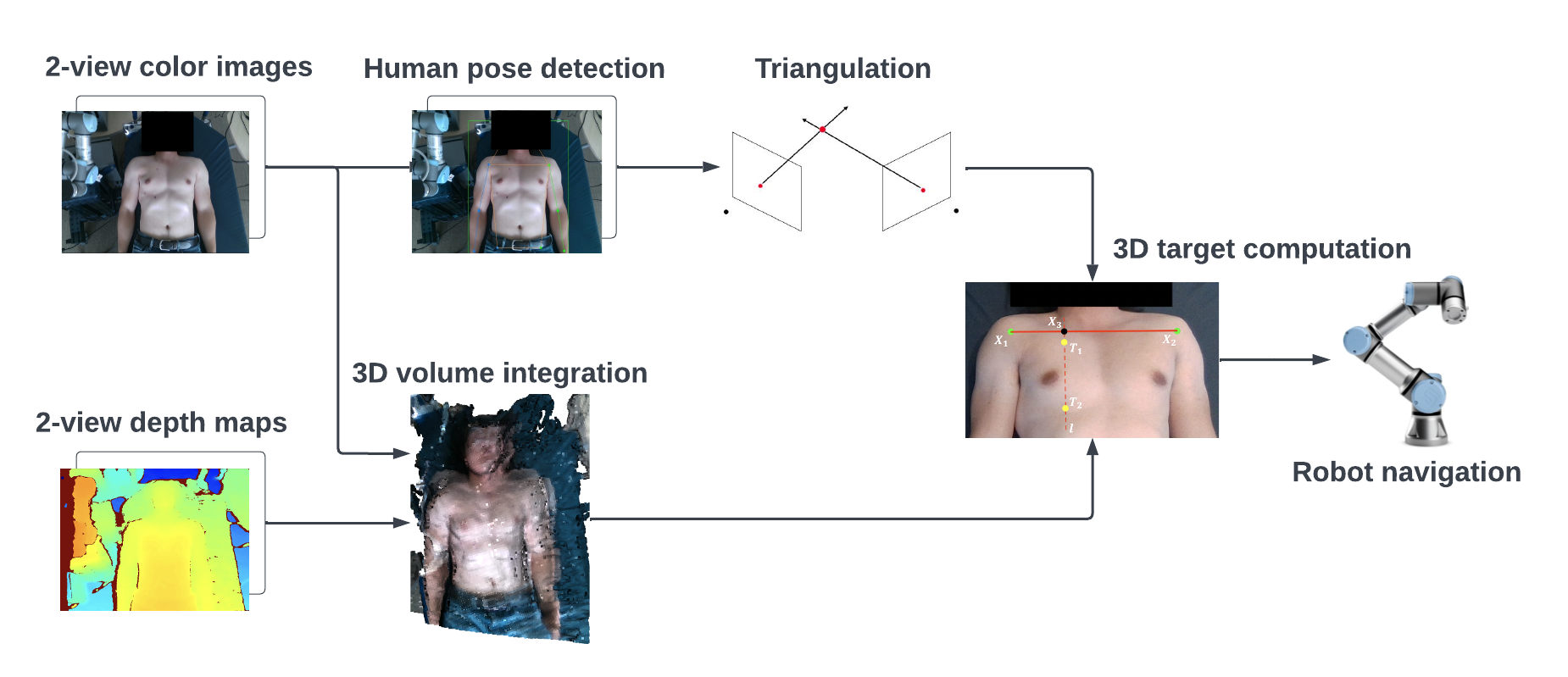}
        \caption{The autonomous scan target localization system pipeline, with images of front targets as example.}
        \label{Pipeline}
    \end{subfigure}
    \hfill%
    \begin{subfigure}[b]{0.32\textwidth}
        \centering
        \includegraphics[width=\textwidth]{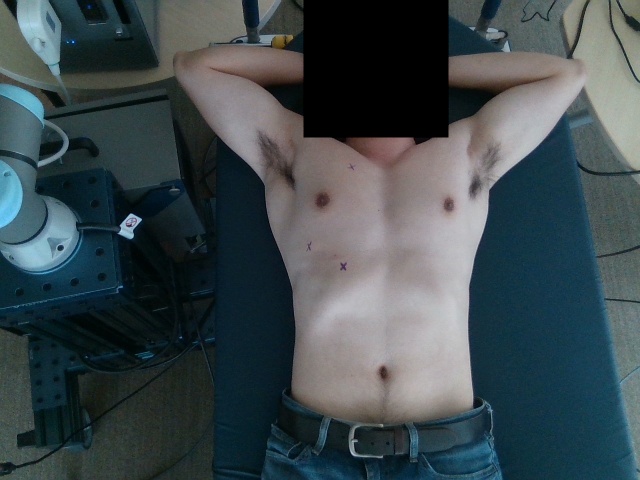}
        \caption{Example image for localizing the side target.}
        \label{sidepose}
    \end{subfigure}
    \caption{The autonomous scan target localization system pipeline. Two-view color images are required for pose estimation and triangulation, while 3D volume integration requires two-view depth maps in addition to that. Target computation uses the outputs from both branches of the pipeline. This procedure is repeated for both front and side targets, with the subject maintaining different body poses.}
\end{figure*}

Fig. \ref{Pipeline} illustrates the general pipeline of our method. First, two-view color and depth images of the subject lying on the stretcher are captured from the two cameras. A pose estimation model is used to extract the 2D keypoint features of the subject. On the other hand, a 3D point cloud is constructed by 3D volume integration using the two-view RGB-D images, which also yields the surface normal vector estimation of all the points. We leverage the 2D and 3D information to compute the coordinates of the target scan points and their surface normal vectors in the global frame. Specifically, triangulation algorithm is applied to compute the 3D global coordinates of keypoint features present in both color images, which are then used to regress the target scan locations. This process is performed twice, for scan locations on the front (AAx) and side (MC) of the body, for which the subject is asked to maintain different postures during the experiment (Fig. \ref{sidepose}). 


\subsection{Hand-eye Calibration}
We utilize hand-eye calibration to obtain the transformation matrix from the camera coordinate frame to the robot base coordinate frame, with the latter referred as the global frame. This $4 \times 4$ transformation matrix (rotation and translation) is denoted as $T_{cam}^{base}$. Since both cameras are stationary with respect to the robot base, we adopt the eye-to-hand calibration setting. A calibration pattern is fixed on the robot gripper and the robot is programmed to move across the field of view of the camera, while color images capturing the calibration pattern are recorded along with the corresponding robot gripper poses. Specifically, we use Apriltag \cite{olson2011apriltag} as the calibration pattern.

\begin{figure}[ht]
    \centering
    \includegraphics[width=6.5cm]{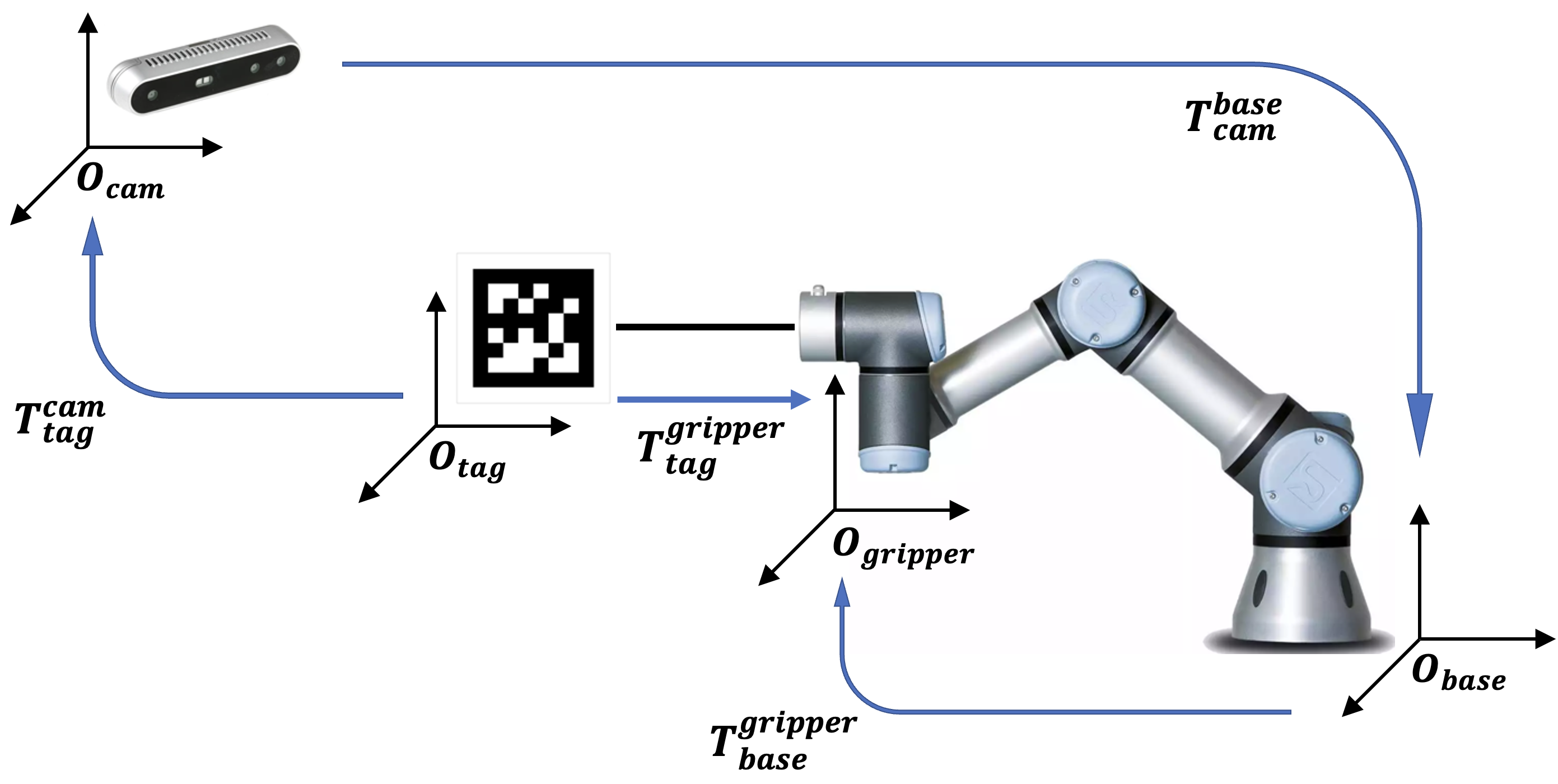}
    \caption{Coordinate frame transformation in eye-to-hand calibration, including robot base, robot gripper, Apriltag and camera. The arrows indicate the direction of the transformation.}
    \label{Eye-to-hand}
\end{figure}

Overall, there are four transformation matrices involved in the calibration process, between robot base, camera, Apriltag and robot gripper. The relationship between these transformations is shown in Fig. \ref{Eye-to-hand}. We implement the $AX=XB$ solver proposed by Park et al. \cite{park_robot_1994} to obtain the estimation for the target transformations. In practice, data from more than 20 different poses were collected for each camera.

\subsection{Pose Estimation}
A deep learning based human pose estimation (HPE) model is deployed to extract keypoint features from the two-view color images of the subject. For our task, we need accurate estimation for the locations of the two shoulders and the right hip of the subject in each image. To evaluate the effect of different pose estimation models on the overall system performance, we compare the performance using two Vision-Transformer (ViT) based models and one Convolutional-Neural-Network (CNN) based model. 

We use ViTPose \cite{xu2022vitpose}, the current state-of-the-art pose estimation model, in our pipeline. It follows the common top-down setting, with a vision transformer as encoder backbone. To speed up the runtime, a YOLOv3 model \cite{redmon2018yolov3} is used as person detector. 

We also test our pipeline with the OpenPose model \cite{cao2017realtime}, which adopts the bottom-up approach and can achieve real-time multiperson pose estimation. First the model produces confidence maps that show the probabilities of a pixel being specific anatomical keypoints, as well as a vector field that indicates the direction of the limb extension. This information is then used to group individual skeletons together.

The performance of the HPE models used in our experiments on the Microsoft Common Object in Context (MS COCO) \cite{lin2014microsoft} validation set is summarized in Table \ref{HPE models}.

\begin{table}[h]
\centering
\scalebox{0.9}{
\begin{tabular}{|c|c|c|c|}
\hline
Model     & Approach  & Train Dataset           & AP   \\ \hline
OpenPose  & Bottom-Up & COCO                    & 61.8 \\ \hline
ViTPose-B & Top-Down  & COCO                    & 75.8 \\ \hline
ViTPose-L & Top-Down  & \makecell{COCO+AiC+\\MPII+CrowdPose\\\cite{lin2014microsoft,wu2017ai,andriluka20142d,li2019crowdpose}} & 79.1 \\ \hline
\end{tabular}
}
\caption{Summary of HPE models tested in our experiments, including the design approach, train dataset and results on MS COCO validation set. AiC: Attributes in Crowd Dataset, MPII: Max Planck Institut Informatik Human Pose Dataset. AP: average precision, a common metric for evaluating HPE model performance.}
\label{HPE models}
\end{table}

\subsection{Triangulation}
Triangulation is a classic computer vision technique used for estimating 3D coordinates from corresponding 2D pixel coordinates in multiview geometry \cite{hartley1997triangulation}. Our goal is to compute the 3D coordinates of three keypoints (two shoulders and the right hip) present in the two-view images. We utilize provided functionality by the OpenCV library \cite{bradski2000opencv} to solve this problem.

\subsection{3D Volume Integration}
\label{3D_integration}
3D volume integration combines multiview RGB-D images together to obtain a holistic view of the scene. It is used for two purposes, surface normal estimation for computing robot gripper orientation and depth adjustment for triangulation result. We use the volume integration function provided by the \textit{Open3D} library \cite{Zhou2018}. Besides the RGB-D data, the intrinsic and extrinsic information of the cameras are required as inputs. The output can be represented as point cloud format, from which the global coordinate and surface normal of each 3D scene point can be extracted.

The surface normal of the target point is estimated with a nearest neighbor method. We assume that the point cloud is very dense so that the variation of surface normal between the target point and its nearest neighbor in the point cloud is infinitesimal. In practice, we discovered that while the X and Y coordinate estimation from triangulation is accurate, the depth estimation (Z-axis coordinate) is usually slightly deeper than the actual value, which is probably due to the relatively narrow baseline between the two cameras\cite{olson2010wide}. As a result, we conduct the nearest neighbor search in terms of Euclidean distance on width and height dimensions (XY-plane) only. We refer to this metric as the Planar Euclidean Distance. Also, to prevent the robot from driving too deep into the human body, we perform depth adjustment by replacing the estimated depth of target point from triangulation with the depth value of its nearest neighbor in terms of Planar Euclidean Distance. 

Denote the target point as $T$ with 3D coordinate $(X_T,Y_T,Z_T)$, the scene point cloud as $A$, and the nearest neighbor of $T$ in $A$ as $P$, the surface normal $\hat{n}_T$ and depth estimate $Z_T$ of the target can be expressed as

\begin{equation}
\begin{multlined}
    \hat{n}_T = \hat{n}_P \text{, } Z_T = Z_P \\
    \text{where } P = \argmin_{a \in A} ||(X_a, Y_a) - (X_T, Y_T)||_2
\end{multlined}
\label{nn_update_eqn}
\end{equation}

Here, $P$ is the nearest neighbor of the scan target in the point cloud obtained from 3D volume integration, in terms of Planar Euclidean Distance. 

\subsection{Robot Target Pose Computation}
The pose of robot gripper in robot base frame is expressed as a 6D vector $(X,Y,Z,RX,RY,RZ)$, where the first three elements denote the position and the last three denote the orientation as a 3D rotation vector in angle-axis representation.

\subsubsection{Target Position}
\begin{figure}[ht!]
    \centering
    \begin{subfigure}[b]{0.28\textwidth}
        \centering
        \includegraphics[width=5cm]{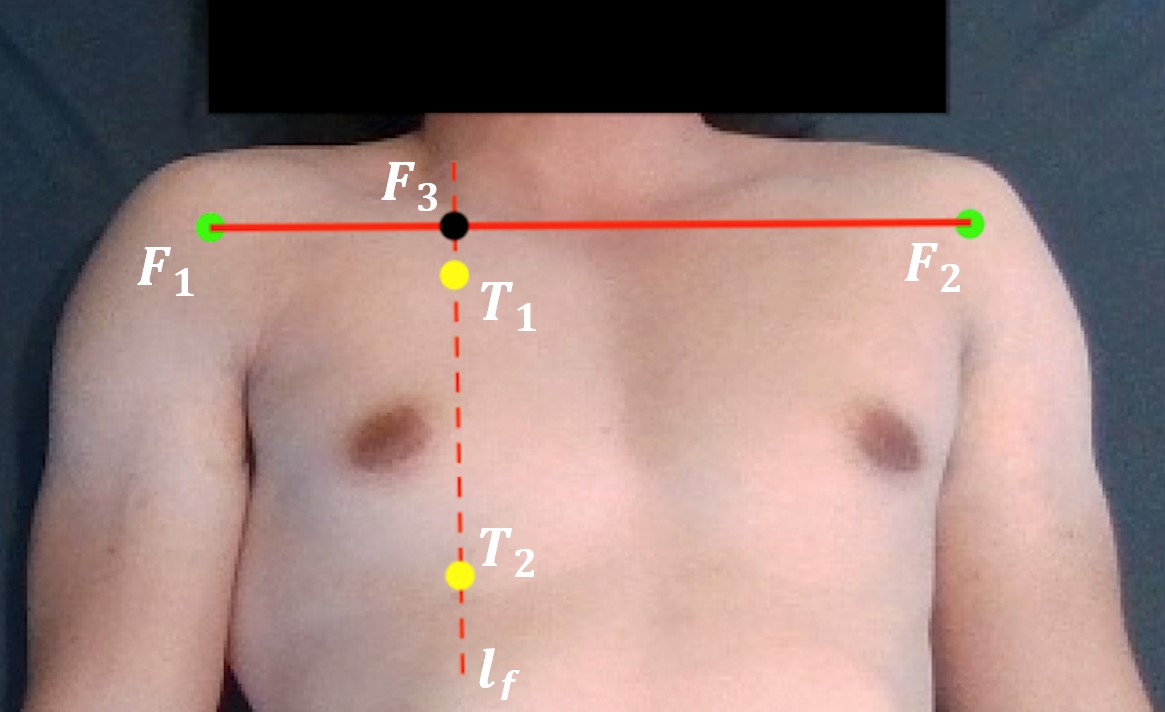}
        \caption{Front pose.}
        \label{front_targets}
    \end{subfigure}
    \hfill%
    \begin{subfigure}[b]{0.18\textwidth}
        \centering
        \includegraphics[width=2.5cm]{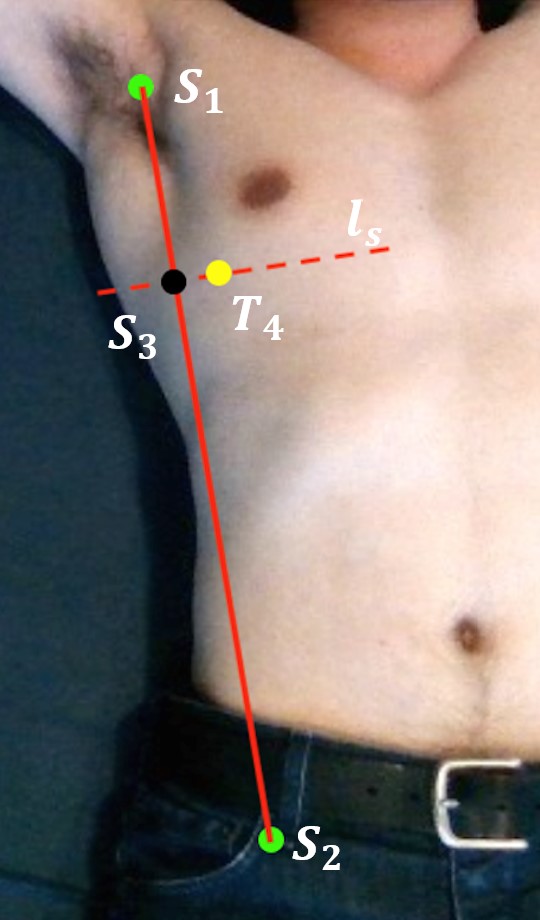}
        \caption{Side pose.}
        \label{side_target}
    \end{subfigure}
    \caption{
        For the front pose, $F_1$, $F_2$ are the 3D coordinates of shoulder keypoints, and $F_3$ is a point on line segment $\overline{\rm F_1F_2}$. The line $\mathit{l_f}$ is perpendicular to $F_1F_2$, passes through $F_3$, and is parallel to the XY-plane of the global frame. The locations of front targets $T_1$, $T_2$ lies on $\mathit{l_f}$. Similarly, for the side pose, $S_1$, $S_2$ are the shoulder and hip keypoints, and target $T_4$ lies on $\mathit{l_s}$.
    }%
    \label{Target position}%
\end{figure}

Due to the scarcity of training data, we use rule-based anatomical models to estimate the three target positions in a 3D setting. We assume identical anatomic proportions for our subjects, a much weaker assumption compared to the constant numerical target UV coordinates adopted by related work \cite{ma2021autonomous}.The regression process is illustrated in Fig. \ref{Target position}.

\paragraph{Front Targets (Fig. \ref{front_targets})}

There are two scan targets located on the front of the human body, namely locations 1 and 2 in Fig. \ref{target_scan_locations}, corresponding to target points $T_1$, $T_2$ in Fig. \ref{front_targets}. We determine these target points from the shoulder locations $F_1$, $F_2$ by introducing the ratio parameters $r_{f1}$, $r_{f2}$. Theoretically, $T_1$, $T_2$ should lie on the mid-clavicular line (MC), which is a line parallel to the anatomic midline \cite{naylor1987midclavicular}. 

Front targets $T_1$, $T_2$ are computed using the same method, the detailed steps are as follows in the case of $T_1$:

\begin{enumerate}
    \item Compute direction vector $\vec{t_1}$ of the line $F_1 F_2$
        \begin{equation}
        \vec{t_1} = \frac{F_2-F_1}{||F_2-F_1||_2}
        \end{equation}
    \item Compute position of $F_3$ on line $F_1 F_2$, $r_{f1}$ is the ratio over line segment $\overline{\rm F_1F_2}$.
        \begin{equation}
        F_3 = F_1 + r_{f1} (F_2-F_1)
        \end{equation}
    \item Compute direction vector $\vec{t_2}$ of the line $\mathit{l_f}$ that is perpendicular to $F_1 F_2$ and parallel to the XY-plane of the global frame. Note that being parallel to a plane is equivalent to being orthogonal to its normal vector, where the normal vector of the XY-plane is just $\vec{n}=\begin{bmatrix}0 & 0 & 1\end{bmatrix}^T$. As a result, $\vec{t_2}$ is the null space of the matrix whose row vectors are $\vec{t_1}$ and $\vec{n}$.
        \begin{equation}
        \vec{t_2} = \mathcal{N}(\begin{bmatrix}
                                    \vec{t_1}^T\\
                                    \vec{n}^T 
                                \end{bmatrix})
        \end{equation}
    \item Compute the position of target $T_1$ on the line $\mathit{l_f} = F_3 + \lambda \vec{t_2}$, where $\lambda$ is an arbitrary scale factor and 
    $r_{f2}$ is the ratio over $\overline{\rm F_1F_2}$. 
    
        \begin{equation}
        T_1 = F_3 + r_{f2} \, \vec{t_2} \, ||F_2-F_1||_2
        \end{equation}
    \item Use \eqref{nn_update_eqn} to set the depth estimate and the surface normal of the 3D target as the depth value and surface normal of its nearest neighbor in the point cloud in terms of Planar Euclidean Distance. The surface normal of the target is used for computing target orientation later. 
\end{enumerate}


With data collected from the subjects, we are able to optimize the ratio parameters $r_{f1}$, $r_{f2}$ of the models. We compute the loss function value with the predicted and ground truth targets, and find the parameters that minimize it. The model for front targets is linear, and its parameters can be optimized using the least square method. 


Since the ground truth 3D target coordinates are also obtained through triangulation algorithm from their pixel coordinates in each view, their depth values could be susceptible to the inherent error of triangulation. As a result, instead of minimizing the 3D Euclidean distance between the predicted and ground truth targets, we minimize the Planar Euclidean Distance between them. The optimization objective could be expressed as:

\begin{equation}
    \label{opt_objective}
    r_{f1}^*, r_{f2}^* = \argmin_{r_{f1} r_{f2} \in \Re} \frac{1}{N} \sum_{1}^{N} ||(X_{Pred}, Y_{Pred}) - (X_{GT}, Y_{GT})||_2 
\end{equation}

Here, $r_{f1}^*, r_{f2}^*$ are the optimized parameters, $(X_{Pred}, Y_{Pred})$ and $(X_{GT}, Y_{GT})$ are the X and Y coordinates of the predicted and ground truth targets respectively.

\paragraph{Side Targets (Fig. \ref{side_target})}

There is one scan target on the side of the human body, which is location 4 in Fig. \ref{target_scan_locations}, and it lies on the Anterior Axillary Line (AAx). It is very difficult to explicitly locate this line on each individual, yet based on empirical observations, we make the assumption that the AAx is parallel to the line connecting the right shoulder and right hip of the subject in 3D space. In this case, the position of right shoulder and right hip are denoted $S_1$, $S_2$ and target $T_4$. The same regression method is used with parameters $r_{s1}$, $r_{s2}$, however with $r_{s2}$ denoting the ratio over $\overline{\rm S_1S_3}$, since the correlation between $\overline{\rm S_1S_2}$ and $\overline{\rm S_3T_4}$ might be weak as $\overline{\rm S_1S_2}$ is too large comparing to $\overline{\rm S_3T_4}$. As a result, model for side target is nonlinear, and we use Stochastic Gradient Descent (SGD) for optimization.


\subsubsection{Target Orientation}

The target orientation can be expressed as the rotation from target frame to base frame $R_{tar}^{base} \in \Re^{3\times3}$, which can be computed as $R_{tar}^{base} = R_{gripper}^{base} R_{tar}^{gripper}$, where $R_{tar}^{gripper}$ is the rotation from the target frame to the gripper frame, aligning the target surface normal to the robot gripper. Here, the Z-axis of the robot gripper should point in the opposite direction of the target surface normal.

\begin{figure}[ht]
    \centering
    \includegraphics[width=6cm]{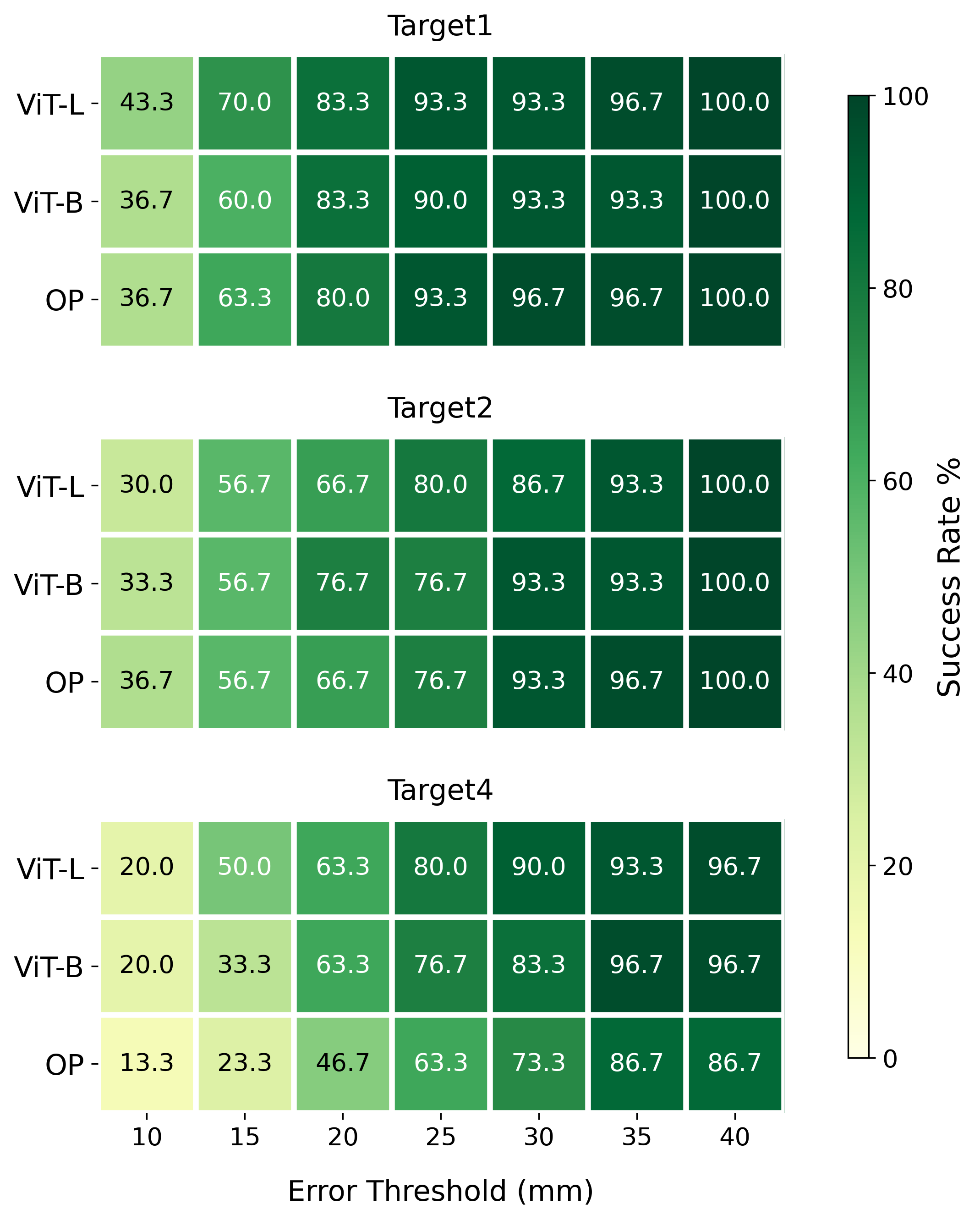}
    \caption{The success rate heatmaps of three targets under increasing error threshold using leave-one-out validation. Faulty results of OpenPose on target 4 are included. ViT-L = ViTPose-Large, ViT-B = ViTPose-Base, OP = OpenPose. Target 4 is the most difficult to localize and target 1 is the least difficult.}
    \label{success rate}
\end{figure}

\section{Experiments\protect\footnote{All subjects participated voluntarily, the Declaration of Helsinki was adequately addressed, and the study was approved by the Institutional Review Board at University of California San Diego (register number 171154)}}
\label{Experiments}

\subsection{Data Collection}
Our subject profile consists of 30 East Asian males with age between 20-27 year-old, height from 168 to 187 cm, and weight between 50 to 130 kg. We use a Butterfly iQ portable US transducer probe to find the target scan points on individual subjects manually. The center point of the contact area between the transducer probe and the subject is marked as the ground truth scan target. 

Each subject is asked to lie on the stretcher with two different poses. For US imaging on the front targets, the arms of the subject stay close to the torso, whereas for side targets the arms are raised with hands behind the head. Two sets of RGB-D images are captured for each subject from the two-view RealSense cameras. The pixel coordinates of the ground truth scan targets in the two-view images are then manually recorded. 


\subsection{Evaluation}

As it is very difficult to directly measure position and orientation errors with physical measurement tools, we conduct offline evaluation after collecting experiment data. With the pixel coordinates of the ground truth scan targets in the two-view images, we derive the 3D coordinates and normal vector using \ref{nn_update_eqn} after triangulation and 3D volume integration. 3D Euclidean distance is used as the metric to evaluate the target position estimation error, while the target orientation error is expressed in terms of the error of normal vector estimation, which is the angle between the predicted and ground truth normal vector. 

In order to achieve robust 3D target localization, we employ two-view RGB-D cameras for US scan target localization. To the best of our knowledge, no other group has investigated the advantages using MVS. To illustrate that our setup yields better 3D reconstruction result than single-view setup, we back-project the computed 3D coordinates of the ground truth scan targets to the corresponding color pictures, and compare the pixel errors of each setting. For single-view setup, we use the provided API in RealSense SDK to deproject a pixel to its 3D coordinate, as in \cite{ma2021autonomous}. In addition, the depth estimate of the 3D coordinate in this setting is updated using the nearest neighbor algorithm for fair comparison.

As the size of our data is small, we use leave-one-out validation strategy. For each metric, we select one sample for validation and the rest for training. The same process is repeated for every sample in the dataset, and the final result is averaged over all the validation samples. 

To accelerate the inference process of deep learning models, we run human pose estimation on GPU (NVIDIA GeForce RTX 3060, 6GB).

\subsection{Analysis}

We observe that the OpenPose model occasionally produces faulty detection results when computing the side target, including the absence and wrong assignment of the right hip keypoint. Given the sample size of 30, OpenPose has two faulty results and the ViTPose models have none. This illustrates that ViTPose is superior than OpenPose in terms of robustness. The faulty results are automatically considered as failures when computing success rate statistics, while samples that lead to faulty results are removed from training and validation sets for other numerical evaluations. 

We define the success of scan target localization as having a target position estimate within a certain error threshold. Fig. \ref{success rate} shows the heatmap of success rates under increasing error thresholds for different HPE models with each scan target. The improvement of HPE model performance has a visible positive impact on the target localization accuracy, especially with low error thresholds and challenging targets. Target 4, locating on the side of the body, is the most challenging one overall. Given an error threshold of 15mm, the success rate of using ViTPose-L is 16.7\% higher than ViTPose-B and 26.7\% higher than OpenPose. 

\begin{figure*}[ht]
    \centering
    \begin{subfigure}[b]{0.48\textwidth}
        \centering
        \includegraphics[width=0.9\textwidth]{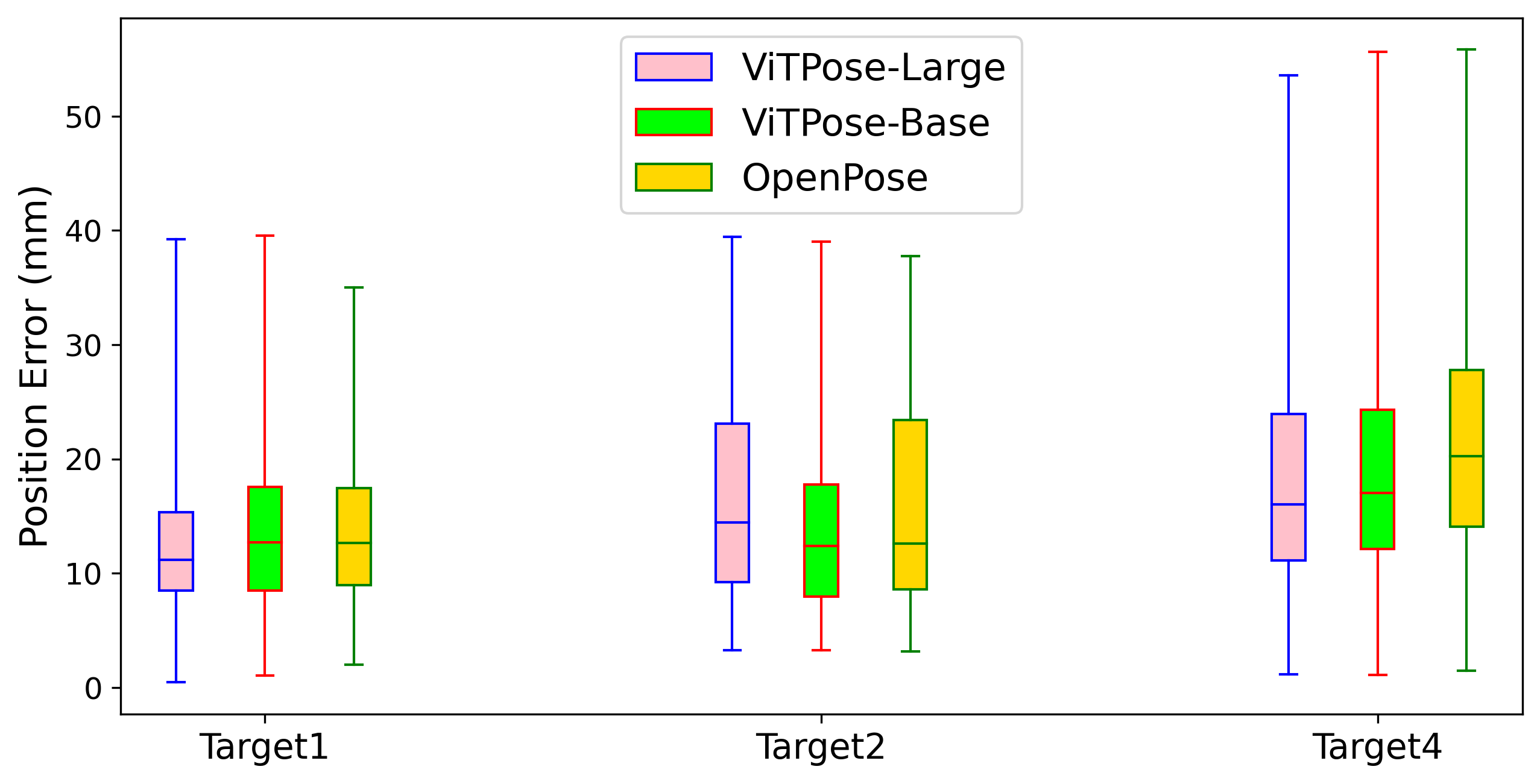}
        \caption{Position error distribution.}
        \label{error dist}
    \end{subfigure}
    \hfill%
    \begin{subfigure}[b]{0.48\textwidth}
        \centering
        \includegraphics[width=0.9\textwidth]{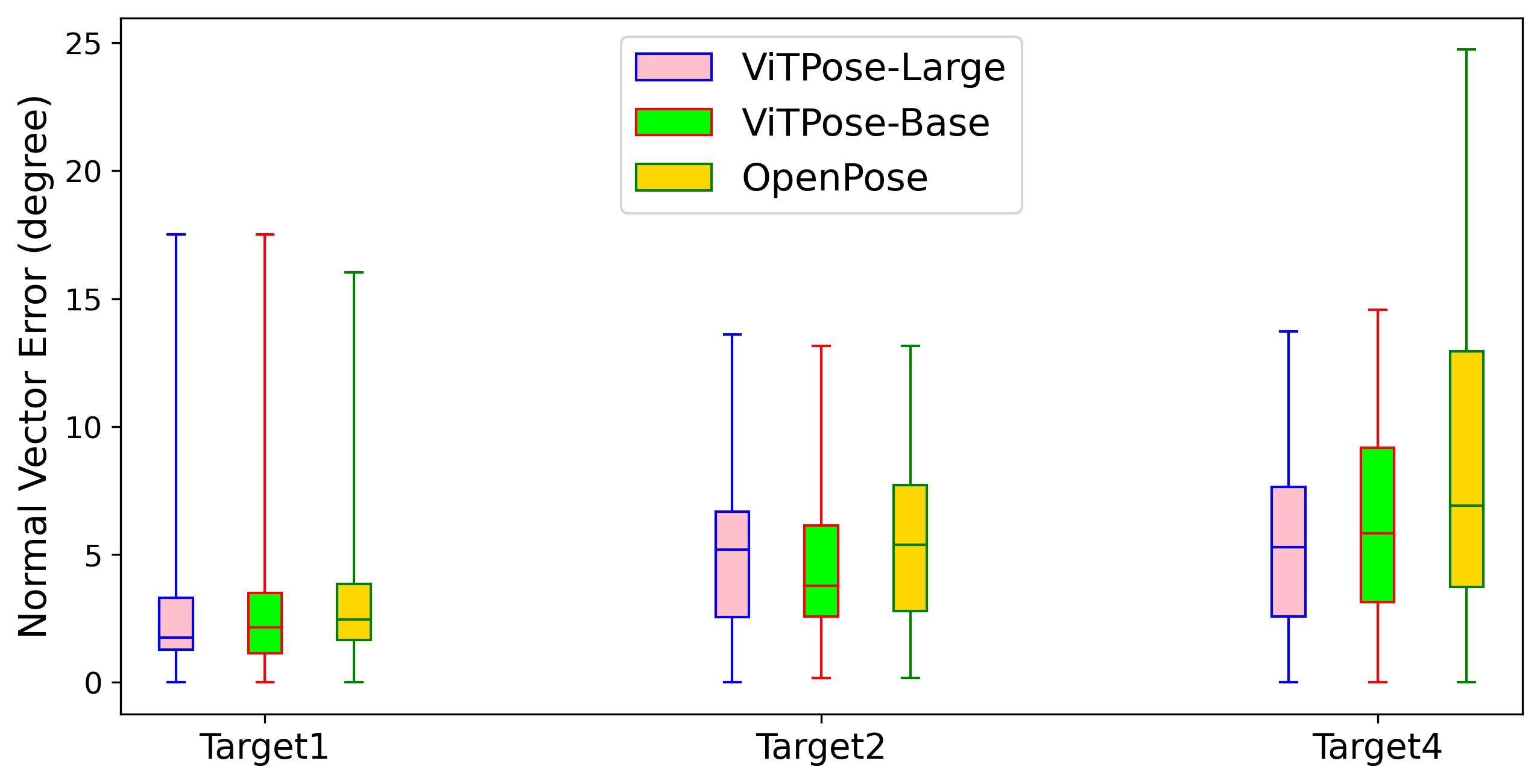}
        \caption{Normal vector error distribution.}
        \label{normal error dist}
    \end{subfigure}
    \caption{Error distribution under leave-one-out validation. Faulty results of OpenPose on target 4 are excluded. The models perform on par with each other for target 1 and 2, but for target 4 ViTPose yields better results than OpenPose.}
\end{figure*}

This conclusion is further corroborated by the absolute error distributions of position and orientation on each target as shown in Fig. \ref{error dist} and Fig. \ref{normal error dist}. All three models have similar performance on targets 1 and 2 in terms of the range and median of error. However, on target 4 they tend to have larger error, especially for the OpenPose model, even with the faulty results excluded. One possible source of error is the deviation between the slopes of the line $S_1S_2$ and the AAx. For OpenPose, the hip keypoint estimation is less consistent in general, and the shoulder estimation degrades significantly with the hands-behind-head body pose for side target scan. Finally, we describe the overall accuracy of our system by the average and standard deviation of errors over all validations using the ViTPose-L model, which is $16.00 \pm 9.79$mm for probe positioning and $4.44 \pm 3.75$\degree for probe orientation.

\begin{figure}[ht]
    \centering
    \includegraphics[width=7.5cm]{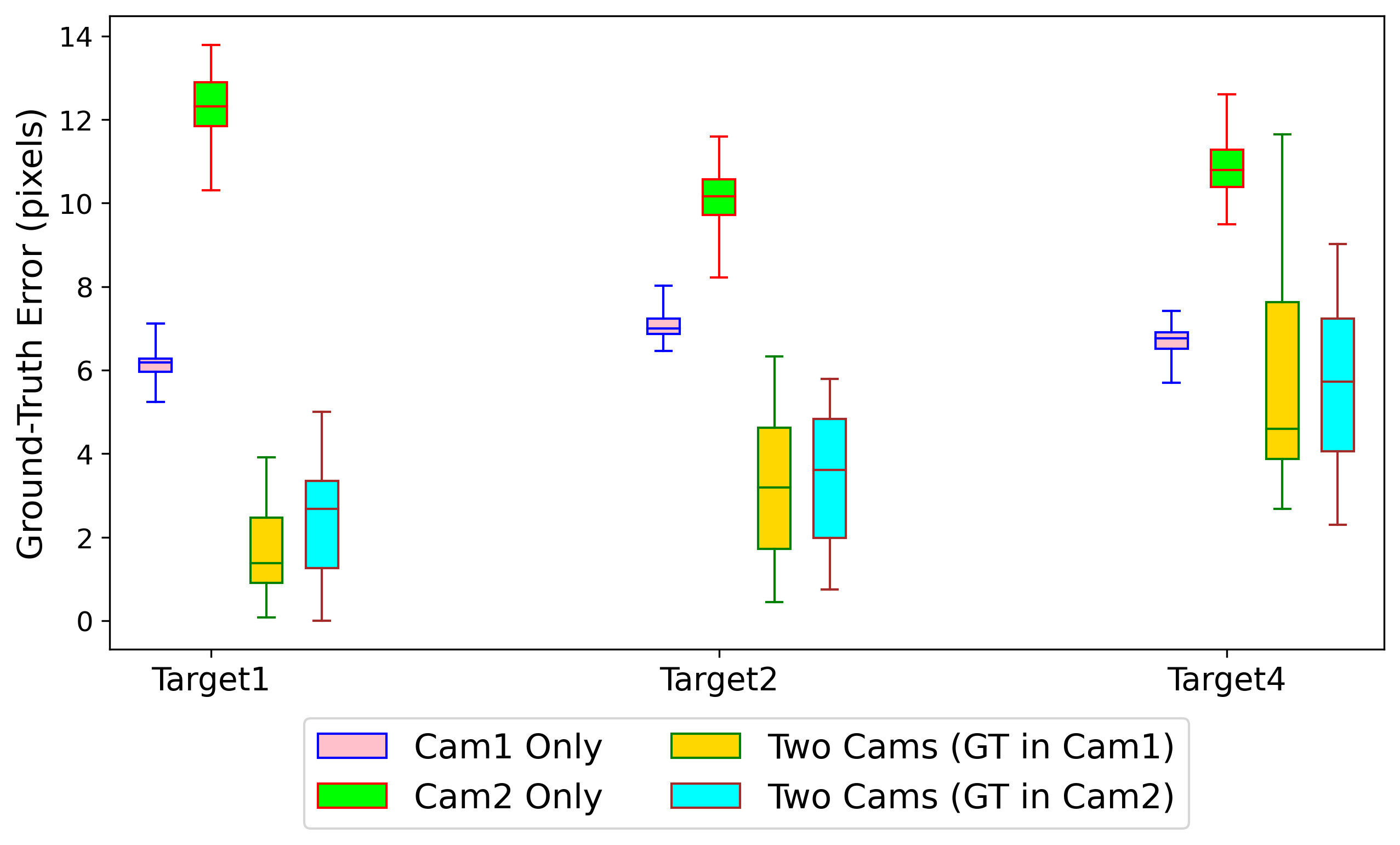}
    \caption{The pixel error distribution of estimated 3D ground-truth target coordinates obtained with two-view and single-view RGB-D camera setups. The 3D coordinates are back-projected to the view images to compute pixel error. Cam1 Only = only using camera 1 in our two-view steup. Two Cams(GT in Cam1) = using two-view cameras and back-projecting the 3D coordinates to the view image of camera 1. Two-view camera setup yields significantly less error across all targets.}
    \label{gt dist}
\end{figure}

Fig. \ref{gt dist} compares the accuracy of ground truth scan target estimation of different camera setups. We compare the pixel errors from using each of the two cameras and using two cameras together. The estimated 3D coordinates are back-projected to the corresponding view images, and the pixel error is computed as the Euclidean distance between the projected pixel coordinate and the ground truth pixel coordinate. The result shows clearly that using two-view RGBD cameras yields significantly less pixel errors across all targets, and is therefore a more reliable option for scan target localization.

\section{Limitation}



The primary limitation of our solution lies in the scan target regression algorithm. By assuming that all human subjects have identical proportions, we are able to learn the relationship between the scan targets and other body part locations without a complex model. However, human proportion can vary among individuals, thus the model could perform suboptimally on very diverse data. Also, the sample size of 30 in our experiment is not large enough to train a model with higher representation power. In the future, given enough data, we can train an end-to-end scan target estimation model by fine-tuning the decoder of the HPE model while keeping the encoder backbone parameters unchanged. 

In addition, we plan to integrate and test our system with other components of autonomous US imaging, including real-time US image processing and force measurement. Future improvement on runtime efficiency is also necessary to address practical problems like body movement during scan.

\section{Conclusion} 
In this paper, we propose a system that addresses the scan target localization problem for lung US imaging. This problem is critical for achieving fully autonomous US scanning of all modalities and yet remains underexplored. As a purely vision based solution, our system adopts a multiview stereo vision setting and incorporates various CV techniques, including human pose estimation, triangulation, and 3D volume integration. We also design a scan target regression model based on explicit assumption on human anatomy. We test our method on 30 human subjects, while related works involves significantly fewer subjects or entirely focus on phantom experiments. The error over all scan targets is $16.00 \pm 9.79$mm for probe positioning and $4.44 \pm 3.75$\degree for probe orientation. For future work, a dataset of larger scale and diversity need to be collected to improve the complexity and generalization ability of the model. Workflow optimization can be done to reduce the runtime of the pipeline. Finally, integration with other components of autonomous US imaging system is necessary to test the overall performance. By succeeding at the challenging task of lung US scan target localization, we believe our approach can be very well generalized to other US imaging modalities that may include procedural US therapies such as US-guided pain injection.

\bibliographystyle{IEEEtran} 
\bibliography{reference}

\end{document}